\documentclass[aps,prl,preprint,groupedaddress]{revtex4-2}
\usepackage{graphicx}
\usepackage{amsmath}

\newcommand{\fluence}{$\rm{J/cm^2}$}

\begin{document}


\title{Mechanism of femtosecond laser ablation revealed by THz emission spectroscopy}


\author{Shuntaro Tani}
\email[e-mail:]{stani@issp.u-tokyo.ac.jp}
\author{Yohei Kobayashi}
\affiliation{1Institute for Solid State Physics, The University of Tokyo, Kashiwa-no-ha 5-1-5, Kashiwa, Chiba, 277-8581, Japan}

\date{\today}

\begin{abstract}
We investigated femtosecond laser ablation dynamics using THz time-domain spectroscopy. To clarify the breakdown dynamics of materials, we focused on the motion of charged particles and measured the terahertz waves emitted during laser ablation. We revealed that the Coulomb force dominated the ablation process. Furthermore, comparisons of the experimental results with theoretical models showed that material breakdown occurs within a few hundred femtoseconds. Our experimental results indicate that electrostatic ablation is the most likely ablation mechanism for metals.
\end{abstract}

\maketitle

When an intense femtosecond laser pulse irradiates a material, the laser field drives the electrons in the material strongly into highly energetic states, which leads to large-scale atomic motions and material ejection. This material removal process is called femtosecond laser ablation and provides a foundation for laser-based machining processes with minimal thermal degradation \cite{Chichkov1996a,Kerse2016}. Extensive research has been conducted into this process for more than three decades, both experimentally and theoretically \cite{Preuss1993,Stampfli1994,Nolte1997,Huang1998,Cavalleri1999,Amoruso2002,Gamaly2011a,Rethfeld2017a,Pompili2018,Takahashi2020,Uchida2020,Tani2022}, to clarify the irreversible processes that cause solids to break down into atoms under application of intense laser fields. Various ablation mechanisms have been proposed and are still subject to debate. These mechanisms include evaporation \cite{Nolte1997}, spallation caused by a laser-induced shock wave \cite{Colombier2007,Winter2020}, explosive boiling \cite{Lorazo2003}, phonon instability \cite{Stampfli1994,Zier2016a}, electrostatic force \cite{Gamaly2002}, and Coulomb explosion \cite{Stoian2002}. As a result, the time required for bond to break varies widely depending on the theoretical model used. For example, \textit{ab initio} molecular dynamics calculations predicted bond breaking times of less than 100 fs \cite{Stampfli1994, Zier2016a}. In contrast, molecular dynamics and hydrodynamics-based calculations predicted bond-breaking times of tens of picoseconds or more \cite{Colombier2005,Lorazo2006,Wu2014,Povarnitsyn2015a}.

Time-resolved spectroscopic studies have revealed the ultrafast dynamics that occur in materials after instantaneous optical excitation. For example, optical pump-probe measurements have revealed the occurrence of changes in both electronic band structures \cite{Huang1998,Callan2000,Winkler2018} and the states of matter \cite{Sokolowski-Tinten1998a,Winter2017}. Additionally, electron beam and X-ray diffraction measurements have shown the energy transfer processes that occur from electronic systems to phonon systems \cite{Sokolowski-Tinten2001,Sokolowski-Tinten2003a,Siwick2003,Waldecker2017,Wang2020a}. However, direct observation of the atomic-level fracture that occurs at several nanometers beneath the surface remains challenging.

One process that will inevitably occur when a solid is broken apart into atoms is the acceleration of the motion of the electrons and ions. These accelerated motions of the charged particles cause emission of electromagnetic waves and these emissions can be measured with subpicosecond time resolution using terahertz time-domain spectroscopy (THz-TDS) \cite{Schmuttenmaer2004,Neu2018}. In particular, measurement of the charged particle motions triggered by laser irradiation, which is called THz emission spectroscopy, has been used for a variety of systems \cite{Beard2002,Sotome2021}. However, although terahertz emission has been reported to occur during laser ablation, the actual ablation mechanism has not yet been revealed \cite{Gopal2013,Liao2016,Jin2016,Tani2017,Tamaki2022}.

In this study, we investigated the ablation mechanism of copper near the material's ablation threshold. We used broadband THz emission spectroscopy to measure the dynamics of the charged particles on a subpicosecond scale. Our vectorial THz-TDS system determined the direction of the charged-particle motion successfully. Furthermore, the charge displacements were estimated from the THz waveforms and compared to theoretical models. As a result, we clarified that ablation occurs within a few hundred femtoseconds and that the electrostatic ablation model is the most plausible of the existing models.
 
 \begin{figure*}
 \includegraphics{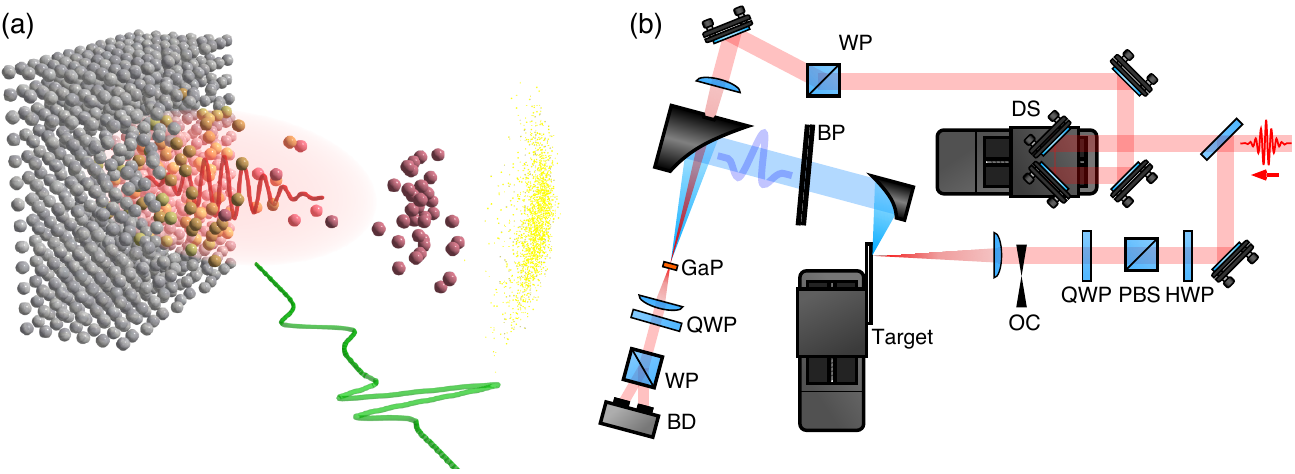}
 \caption{\label{experimental_setup}(a) Schematics of electromagnetic wave emission during femtosecond laser ablation. (b) Experimental setup for THz-TDS of laser ablation. HWP: half-wave plate, QWP: quarter-wave plate, PBS: polarizing beam splitter, OC: optical chopper, BP: black polypropylene film, GaP: gallium phosphide, DS: delay stage, WP: Wollaston prism, BD: balanced detector.}
 \end{figure*}

The experimental setup used for the THz-TDS of the laser ablation process is shown in Fig. 1(b). A Ti-sapphire regenerative amplifier delivered 80-fs pulses that were used as excitation pulses for the ablation process and as sampling pulses for electro-optic detection \cite{Gallot1999}. The laser pulses for ablation were focused perpendicularly onto a target material with a $1/e^2$ spot size of 60 $\rm{\mu m}$. Electromagnetic emissions generated during laser ablation were collected using a pair of off-axis parabolic mirrors on a 310-$\rm{\mu m}$ (110) GaP crystal. Electro-optic sampling with this thin GaP crystal enabled broadband THz measurements to be taken \cite{Leitenstorfer1999,Wu1997}. The parabolic mirror used to for accumulating the THz emissions was placed at an angle of $70^\circ$ to the normal of the target surface. Two black polypropylene films were inserted between the parabolic mirrors to cut out any scattered excitation light while also allowing the THz waves to pass through. Copper plates were used as the target materials. THz-TDS was performed while the sample was moved horizontally on a motorized stage, thus eliminating optical path length variations caused by the ablation process.

Figure 2 shows the electric waveforms measured by the THz-TDS under both linearly and circularly polarized excitation. The field strengths exclude the contribution of Fresnel losses at the black polypropylene films and at the GaP crystal surface. The ablation thresholds for the two polarizations above were 0.2 \fluence\ and 0.3 \fluence\, respectively. Below these thresholds, the electric fields are weak and have a cosine-shaped waveforms. Above these thresholds, the electric fields become much stronger and begin to show a sine-type component. Additionally, a significant polarization dependence is observed at low excitation fluences, whereas this polarization dependence disappears at high excitation fluences.

 \begin{figure}
 \includegraphics{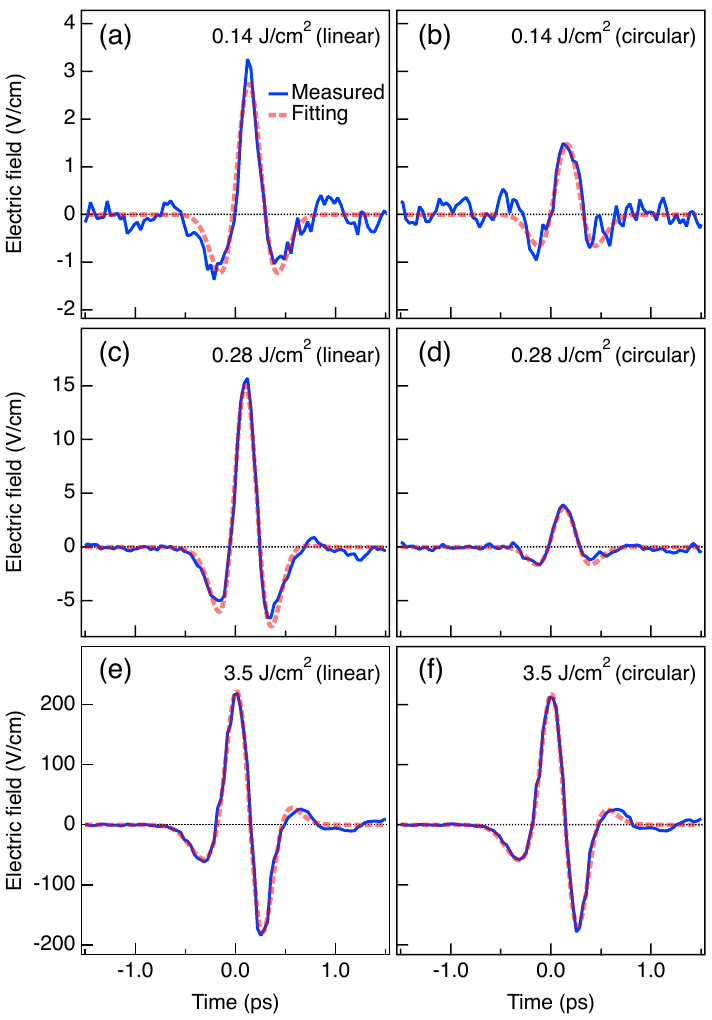}
 \caption{\label{temporal_waveform}Detected THz waveforms for various excitation fluences and polarizations. Copper was used as the target material. The solid lines are the measured data and the dashed lines are the fitting results.}
 \end{figure}
 We confirmed that the observed THz waves originated from the motion of charged particles, based on the dependence of the THz wave polarization on the excitation polarization. We determined the vectorial directions of the THz fields by rotating the nonlinear crystal used to perform the electro-optic sampling \cite{Planken2001}. If the THz wave generation is caused by optical rectification at the target surface, it should be strongly dependent on the excitation polarization \cite{Kadlec2005}. However, the resulting vectorial waveform was linearly polarized along the horizontal plane irrespective of the direction of excitation polarization. We thus conclude that the electromagnetic radiation from the laser ablation process was due to the acceleration or deceleration of charged particles, namely electrons and ions.
 
We can estimate the ultrafast charge motion during laser ablation from the temporal waveform of the THz pulses. The following processes were considered for the estimation: the conversion of transient current to radiation, free-space propagation via the parabolic mirrors, and detection in the electro-optic crystal. Putting all together, we obtained the relationship between the motion of charges and the detected electric field $E_{det}$ (V/cm) \cite{S.Tani}:
\begin{equation}\label{eq:current-to-field}
	E_{det}(t)=-8.0\times{10}^{-10}qN(t)\frac{d^2}{dt^2}v(t),
\end{equation}
where $q$ is the sign of the charge, $N(t)$ is the number of charged particles, and $v(t)$ (nm/ps) is their average velocity. Namely, the detected electric field waveform is the second-order derivative of the transient current.
\begin{figure*}
\includegraphics{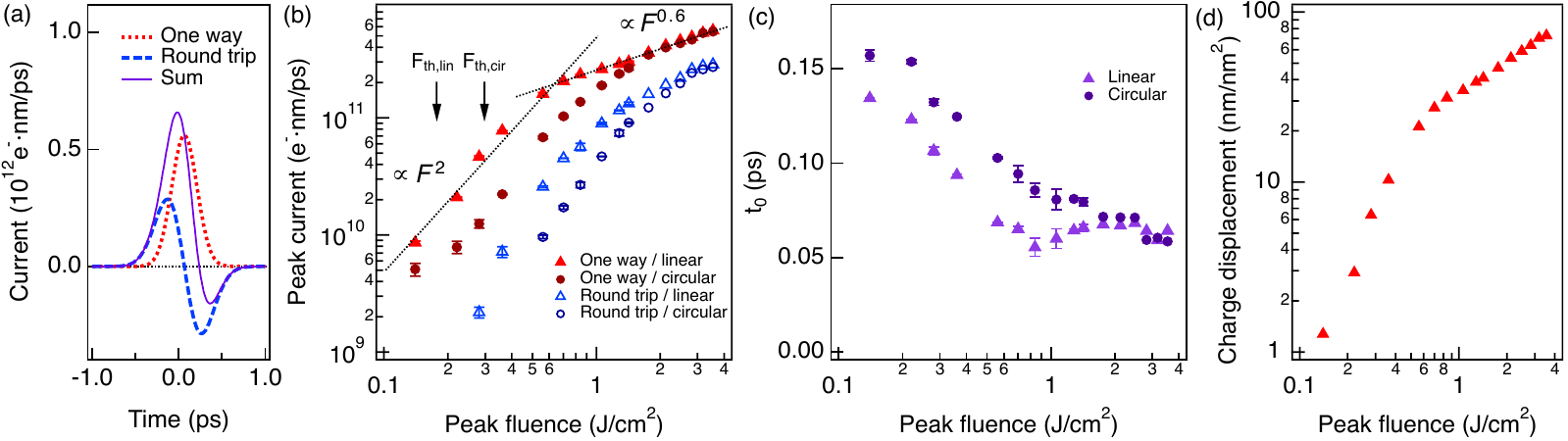}
\caption{\label{fitting_results}(a)One-way and round-trip motion of charges. The excitation fluence was 3.5 \fluence\, and the polarization was linear. (b)Excitation fluence dependence of the magnitude of one-way and round-trip currents. The arrows indicate the ablation thresholds for linear and circular polarization. (c)Fluence dependence of time delay. The time origin is the timing of the excitation pulse irradiation. (d)Charge displacement as a function of fluence for linear polarization.}
\end{figure*}
Here we consider one-way and round-trip motions of the charges. We took a Gaussian and its first-order derivative as the time profile of each component, as shown in Figs.3(a). Under this assumption and Eq. (\ref{eq:current-to-field}), the electric field waveform was fitted with the following waveform:
\begin{equation}
	E(t)=E_2 H_2(\frac{t-t_0}{\tau_2})e^{-\frac{(t-t_0)^2}{\tau_2^2}}+E_3 H_3(\frac{t-t_0}{\tau_3})e^{-\frac{(t-t_0)^2}{\tau_3^2}},
\end{equation}
where $H_n$ is the $n$th-order Hermite polynomial, $E_n$ and $\tau_n$ are fitting parameters representing the amplitudes and timescales, and $t_0$ is the time delay of both components. This fitting well reproduces the experimental results for all excitation fluences and polarizations, as shown in Fig. 2. The values of $\tau_2$ and $\tau_3$ are approximately 0.22 ps and 0.28 ps, respectively, and are nearly constant irrespective of excitation fluence and polarization.

Figure 3(b) shows the excitation fluence dependence of the peak values of the one-way and round-trip transient currents. For both currents, there is a large polarization dependence at fluences below 1.0 \fluence. In contrast, above 1.5 \fluence\, the polarization dependence almost disappears. The one-way currents increase according to a power law, with an exponent of 2 at the low fluence side and 0.6 on the high fluence side.

Figure 3(c) shows the fluence dependence of the common time delay. It decreases monotonically as fluence increases and becomes almost constant above 1.0 \fluence. The time origin was determined by the sum frequency generation of the sampling pulse and the scattered light at the sample surface of the excitation pulse. The optical path length difference due to the black polypropylene films and the temporal walk-off in the nonlinear crystal were compensated.

In this letter, we focused on the displacement of charges by integrating the motion. The displacement of charge density per unit area $\delta$ is:
\begin{equation}
	\delta = \int \Bigl[n_e(t)v_e(t) - n_i(t)v_i(t)\Bigr]dt,
\end{equation}
where $n(t)$ is charge density, and $v(t)$ is the average velocity of the particles. The subscripts $e$ and $i$ represent electrons and ions, respectively. The results are shown in Fig. 3(d). The product $n(t)v(t)$ corresponds to the sum of the one-way and round-trip motions, as per Fig. 3(a). Note that the integral of the round-trip motion is zero, and the magnitude of the final displacement does not depend on a way of decomposing the charge motions.

Let us first assume that the observed motions are only due to electrons. The one-way motion results in the electrons resting on a finite displacement from the surface. A force is needed to slow down the emitted electrons and keep them outside the material. This force can only be the Coulomb force between the electrons and the mirror charges on the metal surface. 
\begin{figure}
\includegraphics{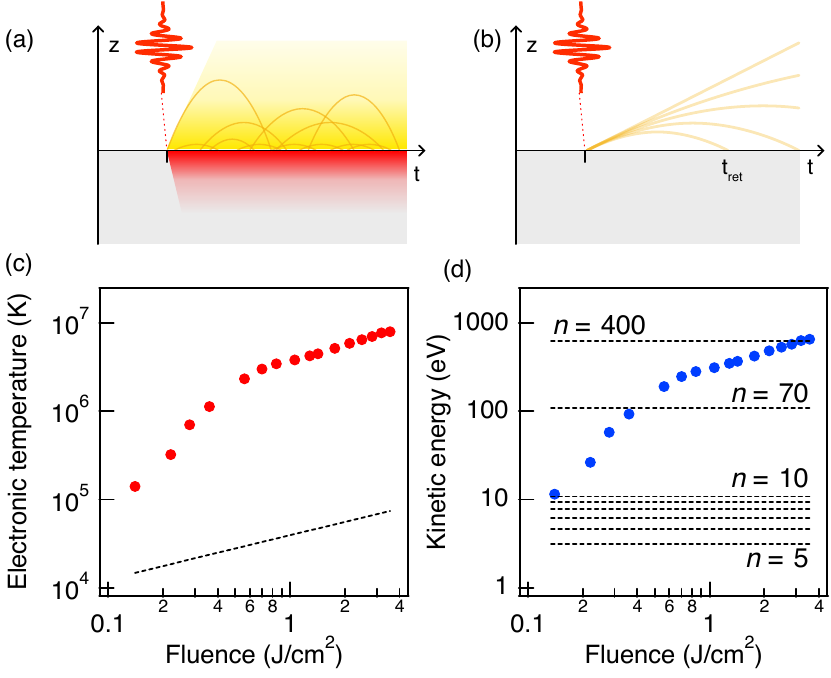}
\caption{\label{electron_spreading}(a)Continuous thermionic emission model. (b)Impulsive photoemission model. (c)Electron temperature at the material surface estimated from the equilibrium state model. The dash line is the electron temperature estimated from the optical penetration length and electron specific heat of copper. The reflectivity of copper is assumed to be 80\%. (d)Kinetic energy per electron required in the impulsive model. The dashed lines are the number of photons required for multiphoton photoemission.}
\end{figure}
Two opposite scenarios can cause such displacement: continuous thermionic emission and impulsive photoemission. In the thermionic emission scenario, the electron system heated by a laser pulse continues to emit thermal electrons, as per Fig. 4(a). In this situation, thermionic emission and the Coulomb force balance and create outward displacement. We derived the electron distribution $\rho_e(z)$ in a self-consistent manner, assuming thermionic electron emission by the Richardson-Dushman formula \cite{Dushman1930} and an exponential spread of the emitted electrons. The distribution is as follows:
\begin{eqnarray}
	\rho_e(z) &=& \rho_s e^{-\beta z}\ \ (z > 0), \\
	\rho_s &=& \frac{m_e k_B T_e}{2\pi^2\hbar^3}\sqrt{\frac{\pi m_e k_B T_e}{2}}e^{-\frac{W}{k_B T_e}}, \\
	\beta &=& \sqrt{\frac{e^2 \rho_s}{\epsilon_0 k_B T_e}},
\end{eqnarray}
where $\rho_0$ is the density of the spreading electrons, $\beta$ is the inverse of the spreading length, $T_e$ is the electronic temperature at the material surface, $m_e$ is the electron mass, $k_B$ is the Boltzmann constant, $\hbar$ is the Plank constant, $W$ is the work function of copper, $e$ is the elementary charge, and $\epsilon_0$ is the vacuum permittivity. Then, the charge displacement density $\delta$ equals $2\rho_0/\beta^2$. Note that $1/\beta$ is equal to the Debye length. As a result, we found the following relationship:
\begin{equation}\label{eq:thermonic_emission_temperature}
	T_e = \frac{e^2}{2\epsilon_0 k_B}\delta.
\end{equation}
Namely, the electron temperature increase at the surface can be directly estimated by integrating the THz waveform three times with respect to time. Figure 4(c) shows the fluence dependence of the electron temperature at the surface calculated from the THz waveform. The estimated temperature is much higher than the electron temperature estimated from the laser fluence, indicating that this scenario is not valid.

As for the other scenario, let us consider a packet of electrons ejected ballistically by impulsive photoemission processes \cite{Aeschlimann1995,Musumeci2010}, as per Fig. 4(b). The Coulomb interaction with the mirror charge slows down and spreads the packet of electrons. Assuming an infinitely thin electron slab of area density $\sigma_{imp}$ ejected at velocity $v_{imp}$ as the electron packet, we obtain an analytical expression for the electron displacement:
\begin{equation}
	\delta(t) = \left\{
\begin{array}{ll}
\sigma_{imp} v_{imp} t - \frac{e^2 \sigma_{imp}^2}{4 \epsilon_0 m_e}t^2 & (0 \leq t < t_{ret})\\
\frac{\epsilon_0 m_e v_{imp}^2}{e^2} & (t \geq t_{ret})
\end{array}
\right.,
\end{equation}
where $t_{ret} = 2 \epsilon_0 m_e v_{imp}/e^2 \sigma_{imp}$ represents the time required for a portion of the emitted electrons to return to the surface. After $t_{ret}$, the charge displacement is constant irrespective of $t$ and independent of the emitted electron density $\sigma_{imp}$. Thus, this scenario also produces finite outward displacement, and the averaged kinetic energy $E_{kin}$ of the emitted electrons can be calculated directly from the charge displacement:
\begin{equation}\label{eq:photoemission_kinetic_energy}
	E_{kin} = \frac{e^2}{2\epsilon_0}\delta.
\end{equation}
Interestingly, although they are based on completely different assumptions, Eqs. (\ref{eq:thermonic_emission_temperature}) and (\ref{eq:photoemission_kinetic_energy}) have the relationship $k_B T_e = E_{kin}$.
Figure 4(d) shows the kinetic energy $E_{king}$ as a function of fluence. The high kinetic energy cannot be explained by multiphoton photoemission or photoassisted thermionic emission \cite{Musumeci2010}. While electron emission with kinetic energies of a few tens of eV is possible due to above-threshold ionization, an electric field of more than 100 V/nm should be generated on the surface considering the yield of electrons \cite{Aeschlimann1995,Saathoff2008}. An electric field that strong should extract ions and such an ablation mechanism is called electrostatic ablation \cite{Gamaly2002}, as shown in Fig. 5(a).
\begin{figure}
\includegraphics{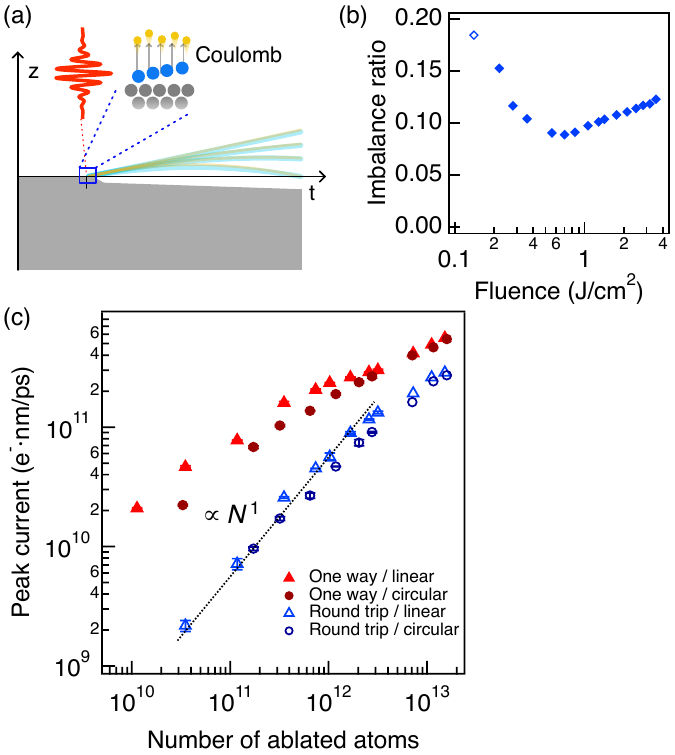}
\caption{(a)Electrostatic ablation. (b)Estimated charge imbalance as a function of fluence. (c)Relationship between transient currents and the number of ablated atoms.}
\end{figure}

Either scenario requires unreasonably high energy electron emission to overcome the strong Coulomb force. Accordingly, our initial assumption that only electrons move on ultrafast timescales must be corrected; if ions move on a subpicosecond time scale, the Coulomb force is suppressed, allowing large displacements to be achieved. Here, we assume the impulsive model and consider an effective particle mixture of electrons and ions. Assuming a mixing ratio of $1/2+f$ to $1/2-f$, the charge displacement yields $\delta = \epsilon_0 E_{kin}/2f^2e^2$. Assuming that the energy of the incident laser pulse is distributed among electrons and ions, the imbalance $f$ can be estimated as shown in Fig. 5(b). Such a small fraction of ionization has also been reported in time-of-flight measurements on GaAs and silicon excited by femtosecond pulses near the ablation threshold \cite{Cavalleri1999}. 

Our experimental observation revealed that the breakup of solids into atoms should occur on a subpicosecond timescale. This time scale is much faster than the commonly accepted time scales in spallation and evaporation models \cite{Lorazo2006,Wu2014,Povarnitsyn2015a,Winter2017}, which range from several picoseconds to a few nanoseconds. This is because electron-phonon coupling is suppressed inside the solid due to Coulomb shielding, but electrons ejected from the surface pull atoms with bare Coulomb. Gamaly et al. estimated the time scale of breakup by electrostatic ablation to be as fast as tens of femtoseconds \cite{Gamaly2011a}, which is consistent with our experiments. Furthermore, the ultrafast breakdown is reasonable to explain the excitation polarization dependence observed in this study and the gentle ablation where the optical penetration length determines the ablation depth \cite{Nolte1997,Hashida2002,Tani2018}.

Finally, Figure 5(c) plots transient current as a function of the number of atoms removed by laser ablation. The number of ablated atoms was obtained by measuring a depth profile of the grooves created by the ablation using a three-dimensional microscope. Remarkably, the round-trip current is proportional to the number of ablated atoms over two orders of magnitude. The proportionality coefficient is approximately 0.05 $e\cdot\rm{nm/ps}$, irrespective of the excitation polarization. This relationship suggests that round-trip current is closely related to laser ablation.

In conclusion, we measured the motion of charged particles during laser ablation in the vicinity of the ablation threshold. The calculated charge displacements reveal that the motion of the ions must occur on a time scale of a few hundred femtoseconds. Our result clarifies the ablation mechanism and its timescale, a matter of long-standing controversy. Moreover, a round-trip current has been found that is proportional to the number of ablated atoms over two orders of magnitude. Precise measurement of ultrafast charge transfer will open the physics of irreversible processes driven by ultrashort laser pulses.

\begin{acknowledgments}
This work was supported in part by New Energy and Industrial Technology Development Organization (TACMI project, P16011); Council for Science, Technology and Innovation; Cabinet Office, Government of Japan; National Institutes for Quantum and Radiological Science and Technology (Cross-ministerial Strategic Innovation Promotion Program); Ministry of Education, Culture, Sports, Science and Technology (JPMXS0118067246 and JPMJCE1313) ; JSPS KAKENHI Grant Number JP20K15188. 
\end{acknowledgments}


\end{document}